# Effect of anti-crossings with cladding resonances on ultrafast nonlinear dynamics in gas-filled PCFs

F. TANI,* F. KÖTTIG, D. NOVOA, R. KEDING, AND P. ST.J. RUSSELL

*Max Planck Institute for the Science of Light, Staudtstrasse 2, 91058 Erlangen, Germany*
*\*Corresponding author: francesco.tani@mpl.mpg.de*

**Spectral anti-crossings between the fundamental guided mode and core wall resonances alter the dispersion in hollow-core anti-resonant-reflection photonic crystal fibers. Here we study the effect of this dispersion change on the nonlinear propagation and dynamics of ultrashort pulses. We find that it causes emission of narrow spectral peaks through a combination of four-wave mixing and dispersive wave emission. We further investigate the influence of the anti-crossings on nonlinear pulse propagation and show that their impact can be minimized by adjusting the core-wall thickness in such a way that the anti-crossings lie spectrally distant from the pump wavelength.**

Gas-filled capillary fibers have been extensively used for nonlinear optics experiments over the last half-century [1]. Nowadays, they are routinely used for spectral broadening and temporal compression of ultrashort mJ-level laser pulses [2], frequency conversion [3] and even for phase-matched high-harmonic generation to the x-ray spectral region [4]. The last 15 years have seen the emergence of hollow-core photonic crystal fibers that guide by anti-resonant reflection (ARR-PCFs). ARR-PCFs come in two main varieties: kagomé-style [5] and single-ring (SR) [6-8]. In contrast to wide-bore capillaries, ARR-PCFs provide low-loss broadband guidance even for core diameters of tens of microns, reducing the pulse energies required for strong nonlinear light-matter interactions from the mJ to the μJ level and allowing scaling from kHz to MHz repetition rates [9–11]. In particular, soliton-related and even strong-field dynamics can be accessed in gas-filled ARR-PCFs pumped with near-infrared ultrashort laser pulses owing to their weak anomalous dispersion across the visible and infrared regions for a wide range of gas pressures [12, 14]. As a result, ARR-PCFs provide a convenient platform for studies of nonlinear dynamics as well as for applications such as ultrashort pulse compression to the single-cycle limit [15], efficient generation of broadband deep and vacuum ultraviolet (UV) radiation both via dispersive wave (DW) generation and four-wave mixing (FWM) [11, 16–19], and impulsive excitation of Raman coherence for multi-octave supercontinuum generation [20]. All these processes rely not only on the broadband guidance offered by ARR-PCFs, but crucially on their weak, spectrally flat dispersion, which can be accurately approximated using a simple hollow-capillary model [21, 22]. In reality, the broad transmission bands exhibited by kagomé-PCF and SR-PCF are interrupted by localized high-loss regions caused by anti-crossings between the core mode and resonances in either the core walls for kagomé-style PCFs or in the capillary walls for SR-PCFs. The wavelength of the $q$-th order anti-crossing can be predicted to good accuracy by the expression $\lambda_q = 2\pi c / \omega_q = (2t/q)(n^2 - 1)^{1/2}$, where $c$ is the speed of light in vacuum, $t$ is the wall thickness, $n$ the refractive index of the glass and $q$ an integer number [23, 24].

Here we study the effects of these anti-crossings on the nonlinear propagation of ultrashort laser pulses in gas-filled ARR-PCFs. We find that the strong, spectrally localized, modification of the real part of the modal effective refractive index creates new phase-matching routes for parametric nonlinear processes such as FWM and DW generation, leading to the emission of strong narrowband light peaks localized in the spectral vicinity of the resonances, despite the high confinement loss. In addition, we investigate how soliton self-compression and ultraviolet DW generation are impaired when the anti-crossings lie close to the pump wavelength, and demonstrate that by appropriate optimization of the fiber structure, in particular tailoring the core-wall thickness, these detrimental effects can be mitigated to great extent.

In the spectral vicinity of an anti-crossing, the capillary model is no longer valid. The influence of the resonance on the real part of the effective refractive index can however be captured by introducing a phenomenological Lorentzian response function $R_q$ centered at the anti-crossing frequency $\omega_q$ [25]. This allows us to incorporate the anti-crossing dispersion for the resonances into the capillary model :

$$n_{\text{eff}}(\omega) \approx 1 + \left[ n_{\text{gas}}^2(\omega) - 1 - \left(2cu_{mn}/(\omega d)\right)^2 + \sum_q R_q(\omega) \right]/2,$$
$$R_q(\omega) = A_q^2 / (\omega_q^2 - \omega^2 + i\Gamma_q \omega) \quad (1)$$

where $n_{gas}$ is the refractive index of the gas filling the fiber core, $u_{mn} = 2.405$ for the fundamental core mode, $\omega$ is the frequency, $d$ is the core diameter and the $R_q$'s are zero in the anti-crossing-free case (orange-dashed line in Fig. 1). The amplitude $A_q$ and damping $\Gamma_q$ of the $q$-th resonance are estimated by fitting the modal index $n_{eff}$ calculated through FEM to Eq. (1). In general, both coefficients depend on the precise fiber structure and in particular on the uniformity of the core-wall thickness: the larger the non-uniformity, the broader the resonance. The approximation used to derive Eq. (1) allows us conveniently to add the Lorentzian contributions to the dispersion.

Figure 1 shows the wavelength-dependent effective refractive index (blue dots) of a kagomé-PCF with $d = 46$ µm and $t = 286$ nm, and a SR-PCF with $d = 53$ µm and $t = 350$ nm, calculated by finite-element modeling (FEM) in the vicinity of the $q = 1$ resonance. These fibers were used in the experiments; their scanning electron micrographs are shown as insets in Fig. 1.

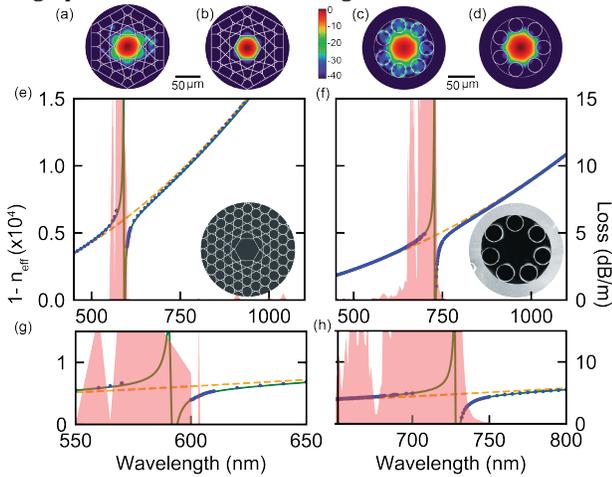

**Fig. 1.** Effective refractive index of the fundamental core mode of an ideal kagomé-PCF (e) and an SR-PCF (f), calculated using FEM (blue dots), and with the capillary model given by Eq. (1) with $R_1 = 0$ (orange-dashed line). Fitting the FEM data to Eq. (1) including the response function given by Eq. (1) (green-solid line) gives $A_1 = 3.5$ THz, $\Gamma_1 = 10$ THz for the kagomé-PCF (left) and $A_1 = 2.4$ THz, $\Gamma_1 = 4.1$ THz for the SR-PCF. (g, h) Zoom into the anti-crossing. Both fibers are assumed to be evacuated. The red under-shaded trace shows the FEM-calculated fiber loss. Scanning electron micrographs of the fibers are shown as insets. The plots above show the FEM mode profiles at 1030 nm (off-resonance) and at 610 nm and 700 nm (on-resonance) for the kagomé-PCF (left) and SR-PCF (right).

Predictions of Eq. (1) are in excellent agreement with the FEM data obtained for the two fiber structures, both in the vicinity of and far away from resonances (Fig. 1, solid-green curve). Note that the non-circular core boundary increases the effective core diameter of the SR-PCF [26], so that the best fit of Eq. (1) to the FEM data occurs for a core diameter $d \sim 6\%$ larger than the distance between two diametrically opposite capillaries.

Because of the anti-crossing, wavelengths on opposite sides of the resonance can have the same effective refractive indices. Through nonlinearity, this can lead to an energy transfer between these wavelengths provided the overall photon energy is conserved. When the fiber is filled with a noble gas, the nonlinearity is given mainly by the optical Kerr effect. As a result, new spectral components can be generated or amplified in the vicinity of the resonance via FWM or DW emission, which can also be described as cascaded FWM [27]. In the case of degenerate FWM, two pump photons at frequency $\omega_0$ generate an idler photon at frequency $\omega_i$ and a signal photon at frequency $\omega_s$. For this interaction, energy and momentum conservation require $\Delta\omega = 2\omega_0 - \omega_s - \omega_i = 0$ and $\Delta\beta_{FWM}(\omega) = \beta(\omega_s) + \beta(\omega_i) - 2\beta(\omega_0) + \Delta\beta_{Kerr} = 0$, where $\beta = n_{eff}\omega/c$ is the modal propagation constant, $\Delta\beta_{FWM}$ is the FWM dephasing rate and $\Delta\beta_{Kerr} = 2\gamma P_0$ is the nonlinear correction due to the optical Kerr effect ($\gamma$ is the nonlinear fiber parameter and $P_0$ the pump peak power)[26]. Figure 2(a) plots $\Delta\beta_{FWM}$ for the kagomé-PCF when filled with 35 bar Ne and pumped with pulses of peak power $P_0 = 450$ MW (the parameters is obtained from the numerical simulation corresponding to the experiment described later). Under these conditions the anti-crossing-related refractive index change strongly alters the landscape of allowed FWM interactions, introducing new narrowband solutions indicated by the curves originating at ~600 nm, where phase-matching is satisfied for many different combinations of pump and idler frequency. As a result, a pulse with a spectrum broad enough to contain the pump and idler frequencies will transfer part of its energy to the signal in the vicinity of $\omega_q$.

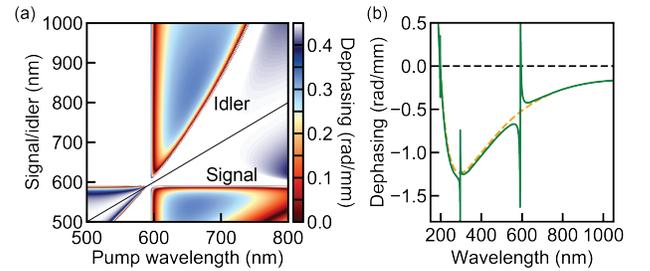

**Fig. 2.** (a) FWM dephasing for the kagomé-PCF, filled with 35 bar Ne. The slanted black line indicates degeneracy (signal and idler frequencies equal to the pump frequency). (b) Dephasing to DWs calculated under the same conditions, for an ideal anti-crossing-free fiber (dashed orange) and the real fiber filled with Ne at 35 bar (green).

The second mechanism—DW emission—can be understood as light shed by a soliton to a linear wave at a different frequency. Within this picture, the dephasing rate of the interaction is $\Delta\beta_{DW}(\omega) = \beta(\omega) - (\beta_0 + (\omega - \omega_0)\beta_1 + \gamma P_c \omega / \omega_0)$ [28, 29] where $\beta_0$ is the propagation constant and $\beta_1$ the inverse group velocity at the soliton central frequency $\omega_0$ and $P_c \approx N P_0$ is the peak power of the self-compressed $N$-th order soliton [22], with $N = (L_D / L_{NL})^{1/2}$, where $L_D = T_0^2 / |\beta_2|$ is the dispersion length, $L_{NL} = 1/(\gamma P_0)$ the nonlinear length, $T_0$ the initial soliton duration and $\beta_2$ the second-order dispersion. Figure 2(b) plots $\Delta\beta_{DW}$ calculated for the same conditions as the FWM case above, and shows that phase-matching is satisfied in the deep UV at ~200 nm, in the anti-crossing-free case ($R_1 = 0$). Using the effective refractive index in Eq. (1), including higher-order resonances, just as for FWM, the DW dephasing rate is altered by the anti-crossings and an additional DW can be phase-matched at the wavelength corresponding to the $q = 1$ anti-crossing.

As already discussed, anti-crossings that are spectrally far away from the pump is mainly result in emission of narrow spectral peaks in their vicinity, where the loss is high. These peaks are observed in experiments under many different conditions and can be caused by FWM and/or DW emission, depending on the experimental parameters. Figure 3(a) shows the measured output spectrum of a 17-cm-long length of kagomé-PCF, when filled with 35 bar Ne and pumped with 8.4 µJ, 23 fs pulses in the anomalous dispersion region at 1030 nm. During propagation through the fiber, the pulses experience soliton self-compression followed by DW emission at ~206 nm. Narrow peaks are clearly visible in the vicinity of anti-crossings at ~600 nm for the fundamental, and ~300 nm for the first higher-order, core-wall resonances. The amplitude of these peaks was found to be strongly pump-energy-dependent. In particular, we observed a threshold-like behavior, in which the peaks appear only after the tail of the continuum extends to wavelengths shorter than ~600 nm. We attribute the additional spectral peaks at ~600 nm and ~526 nm to inhomogeneities in the thickness of the core-walls in the kagomé-PCF. Similar peaks were observed with several different kagomé and SR-PCFs, their spectral position always coinciding with the anti-crossing wavelengths, irrespective of the specific experimental parameters.

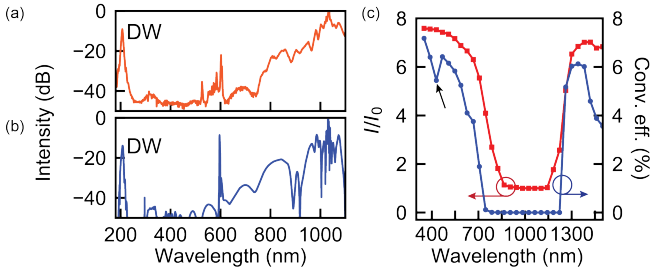

**Fig. 3.** (a) Measured (a) and simulated (b) spectrum at the kagomé-PCF output. (c) Maximum peak intensity along the fiber normalized to the input intensity (red squares) and conversion efficiency to the UV DW (blue dots) as function of the wavelength of the $q=1$ anti-crossing. The arrow points to the drop in the conversion efficiency to the DW, which is caused by the $q=2$ anti-crossing, which coincides with the DW wavelength.

To further confirm the relationship between the narrow spectral peaks observed in the experiments and the anti-crossing-related refractive index change, we performed numerical simulations using a unidirectional field equation [30]. The input pulses were characterized using second harmonic generation frequency-resolved optical gating (SHG FROG) and used in the simulations (the FROG traces are available in a previous paper [11]). The fiber dispersion was modeled using the real part of the effective index in Eq. (1) including higher-order resonances, while the loss was deduced from the imaginary part (in general, this gives higher loss values than FEM). The simulation in Fig. 3(a) agrees well with the experiment in respect of the emission of narrow spectral peaks at the anti-crossing wavelength (~600 nm). These peaks are absent when the fiber is replaced by an anti-crossing-free capillary dispersion model in the simulations.

Depending on their spectral position, the anti-crossings can have a strong impact on nonlinear pulse propagation, impairing effects such as soliton self-compression and DW emission. Using numerical simulations, we investigated this dependence for the case of the kagomé-PCF filled with 35 bar Ne. To be general, we simulated the propagation of a 23 fs (FWHM) Gaussian pulse with 6 µJ energy (this corresponds roughly to the energy contained within the main peak of the pulse used in the experiments) (Fig. 3(b)).

As the $q=1$ anti-crossing approaches the pump wavelength (1030 nm), the maximum peak intensity along the fiber and the conversion efficiency to the UV DW degrade, indicating that soliton self-compression is strongly disrupted. This is because the resonance alters the dispersion landscape, introducing higher-order dispersion that leads to pulse break-up and the formation of side-pulses, which drain energy from the main pulse. As the peak intensity decreases, the shock effect—crucial for efficient UV emission—becomes weaker, resulting in lower conversion efficiency to the UV DW. In contrast, if the anti-crossing is far away from the pump wavelength, it has only a minor impact on the pulse propagation. The small drop in the conversion efficiency to the DW for the $q=1$ anti-crossing at 400 nm (Fig. 3(b)) is caused by the $q=2$ anti-crossing, which coincides with the DW wavelength. Note that the $q=2$ anti-crossing can also impair pulse compression and DW emission when the $q=1$ anti-crossing occurs at wavelengths longer than the pump, leading to a drop in both the maximum peak intensity and the conversion efficiency to the UV DW.

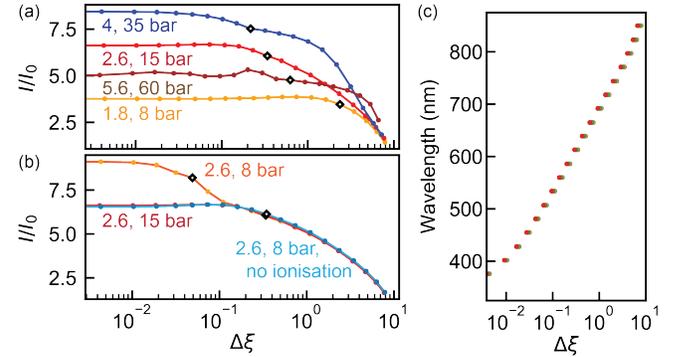

**Fig. 4.** (a) Maximum peak intensity along the fiber as a function of $\Delta\xi$ for increasing $N$ by incrementing the gas-filling pressure (the first number is $N$ and the second the Ne pressure). Each data-point is calculated for a 27 fs (FHWM) Gaussian pump pulse launched into the kagomé-PCF with pulse energy 6 µJ. Each curve is normalized to the peak intensity $I_0$ at the fiber input. The black diamonds mark the points where each $I/I_0$ curve has dropped below 90% of its maximum. (b) The $I/I_0$ curves obtained keeping $N$ constant and changing gas pressure and pulse energy: 6µJ (red curve) and 11.5 µJ (orange curve). (c) Central wavelength of the $q=1$ anti-crossing as a function of $\Delta\xi$ (color-coding is the same as in (a)).

More generally, the impact of the resonances on the pulse propagation dynamics depends on the relative strength of the second- and third-order dispersion and on the pulse bandwidth. In order to quantify this, we introduce the parameter $\xi = L_{D3}/L_D = T_0|\beta_2/\beta_3|$ as the ratio of the second order to the third order dispersion lengths, $L_{D3}/L_D$. In the anomalous dispersion, when the soliton order N>1, the pulse selfcompresses by a factor which is inversely proportional to N over a distance ~LD/N. As a result we expect the impact of the anti-crossings on the soliton compression to increase as $N$ increases. The maximum peak intensity along the fiber and the position of the $q=1$ anti-crossing are plotted in Fig. 4 as functions of $\Delta\xi = \xi_0 - \xi(\lambda_q)$, where $\xi_0$ is calculated using Eq. (1) with $R_1=0$. The soliton order can be tuned either changing the gas pressure (which alters $\gamma$ and

$\beta_2$) or the input pulse energy. As the Ne pressure and *N* increase, the self-compressing soliton reaches shorter durations and higher peak intensities. In the case of weak ionization, the balance between the compression ratio and the quality factor (which goes like $1/N$ for $N>3$ [22]) is optimal for $N\sim 4$, under which conditions the self-compressed pulse reaches maximum peak intensity. As the compression ratio increases, the bandwidth becomes larger and a smaller value of Δξ (i.e., a spectrally more distant anti-crossing) is required for clean and efficient temporal compression (Fig. 4(a)). For soliton order $N \geqslant 4$ the pulse self-compresses over a shorter fiber length and as a result $I/I_0$ begins to drop for larger $\Delta \xi$ (brown curve Fig. 4 (a)).

In the absence of strong ionization, $I/I_0$ is solely a function of Δξ and *N*. To illustrate this, in Fig. 4(b) we plot the $I/I_0$ ratios for two different combinations of gas pressure and pulse energy (6 µJ and 15 bar; 11.5 µJ and 8 bar) chosen so that *N* = 2.6. In the absence of ionization, the two curves coincide perfectly. When however ionization is included, the curves agree only at larger values of Δξ. This is because the additional polarization introduced by free electrons causes stronger temporal compression and a spectral blue-shift [31]. As a result, the pulse wavelength moves closer to the anti-crossing and $I/I_0$ begins to drop at even smaller values of Δξ, as illustrated by the orange curve in Fig. 4(b). Thus, owing to the differing ionization-driven dynamics, the red and orange curves in Fig. 4(b) only coincide at large Δξ where the maximum intensity (and therefore the ionization) is reduced because of the anti-crossing.

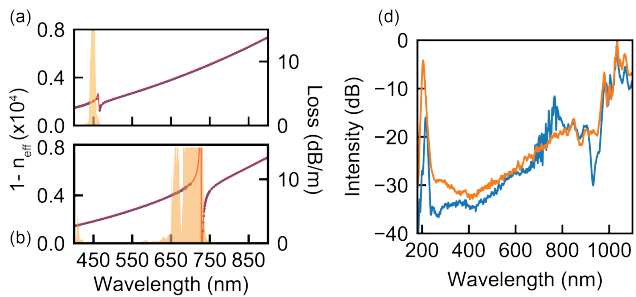

**Fig. 5.** (a) Effective refractive index of the fundamental core mode of the SR-PCF calculated via FEM (blue dots), for the original fiber (∼350 nm capillary-wall thickness); and (b) after etching (∼220 nm). The solid lines show the curves obtained by fitting the FEM data to Eq. (1) and the under-shaded curves plot the FEM-calculated fiber loss. (c) Experimental output spectra at 11.3 µJ input energy in the original fiber (blue) and 9.7 µJ in the etched fiber (orange).

To experimentally investigate the impact of the anti-crossing wavelength, we compared two different fibers: the previous SR-PCF with ∼350 nm capillary wall thickness and the same fiber after being etched with hydrofluoric acid so as to reduce the capillary wall thickness to ∼220 nm. In both cases, the fiber was 15 cm long and filled with 25 bar Ne. Figure 5(a) shows the effective refractive index of the two fibers calculated via FEM. Etching the fiber shifts the $q=1$ anti-crossing from ∼740 to ∼470 nm. In Fig. 5(b), the output spectra of the fiber are shown for input energies of 11.3 µJ (original fiber) and 9.7 µJ (etched fiber). Despite the lower input energy in the etched fiber, the amplitude of the DW at ∼200–220 nm is more than 10 times stronger than in the original fiber, confirming that soliton self-compression is dramatically improved. Additionally, the $q=1$ spectral peaks at ∼740 nm in the unetched fiber shift to ∼470 nm in the etched fiber and become much weaker. Note that the $q=1$ anti-crossing is relatively wide because the capillary-wall thickness varies by ∼40 nm from capillary to capillary.

In conclusion, anti-crossings between the fundamental core mode and resonances in the cladding membranes can strongly affect nonlinear propagation of ultrashort pulses in gas-filled anti-resonant-guiding hollow core PCFs. The change in dispersion introduced by these anti-crossings leads to the emission of light in spectrally narrow bands, through phase-matched FWM and DW emission. The impact of these anti-crossings depends on the soliton order of the pump pulse, and can be minimized by etching the fiber so as to shift them far from the pump wavelength.